\documentclass[preprint,aps,showpacs]{revtex4}
\usepackage[pdftex]{graphicx}
\usepackage{amsmath}
\usepackage{amsxtra}
\usepackage{amstext}
\usepackage{amssymb}
\usepackage{latexsym}
\usepackage{dsfont}
\usepackage{slashed}

\newcommand{\be}{\begin{equation}}
\newcommand{\ee}{\end{equation}}
\newcommand{\bn}{\begin{eqnarray}}
\newcommand{\en}{\end{eqnarray}}

\begin{document}

\title{
Do we have the explanation for the Higgs and Yukawa couplings of the {\em 
standard model}? 
}


\author{Norma Susana Manko\v c  
Bor\v stnik}
\affiliation{Department of Physics, FMF, University of Ljubljana,
Jadranska 19, Ljubljana, 1000}


\begin{abstract}
The {\em  spin-charge-family} theory~\cite{norma,pikanorma,NF,gmdn,gn,NBled2012,HN,NHD} 
offers a possible 
explanation for the assumptions of the {\it standard model}, interpreting the {\em standard 
model} as its low energy effective manifestation~\cite{NF}. The theory predicts several 
scalar fields determining masses and mixing matrices of fermions and weak bosons. The scalar 
fields manifest as doublets with respect to the weak charge, while they are triplets with 
respect to the family quantum numbers. Since free scalar fields (mass eigen states) differ 
from those which couple to $Z_m$ and to $W^{\pm}_{m}$ or to each family member of each 
of the family the {\em  spin-charge-family} theory predictions for LHC might differ from those 
of the {\it standard model}. 
\end{abstract}


\maketitle

\section{Introduction } 
\label{introduction}

The {\it standard model} assumes one scalar field, the Higgs (and the anti-Higgs) with the non zero 
vacuum expectation value and with the weak charge in the fundamental representation. It assumes also 
the Yukawa couplings. The Higgs (and the anti-Higgs), interacting with the weak and hyper gauge 
bosons, determines masses of the weak bosons and, "dressing"  the right handed fermions with 
the needed weak and hyper charges, determines, together with the Yukawa couplings, also the 
masses of the so far observed families of fermions. 

The question is: Where do the Higgs together with the Yukawa couplings  of the {\it standard model} 
originate from? Is the Higgs a scalar field with the fermionic quantum 
numbers in the charge sector? Or there are 
several scalar fields (all doublets with respect to the weak charge), which manifest in the 
observable region as the Higgs and the Yukawa couplings?

To answer any question about the origin of the Higgs and the Yukawa couplings one first needs the 
answer to the question: Where do families originate? And correspondingly: How many 
families do we have at all?

There are several inventive proposals in the 
literature~\cite{jackiw,weinberg,appelquist,hung,shafi,buras,simonov,ryttov,string} 
extending the {\it standard model}. No one explains,  
to my knowledge, the  origin of families.   

There are several proposals in the literature explaining the mass spectrum and mixing matrices of quarks and 
leptons~\cite{fritzsch} and properties of the scalar fields~\cite{Georgi,giudice,belen,ch}. 
All of them just assuming on one or another way the number of families.

I am proposing the theory~\cite{norma,pikanorma,NF,gmdn,gn,AN,HN,NBled2012}~\footnote{Several colleagues, 
first of all is H. Bech Nielsen, and students are participating in this project.},  the 
{\it  spin-charge-family} theory, which does offer the explanation for the origin of families, 
of  vector gauge fields and of several scalar fields: A simple starting action at higher 
dimensions determines at low energies properties of families of fermions, of  vector 
gauge fields and of several scalar fields. The {\it standard model} can be interpreted as a 
low energy manifestation of the {\it spin-charge-family} theory.

Families appear in the theory due to the fact that there are  two kinds of gamma matrices 
(two kinds of the Clifford algebra objects, only two): 
i. One kind  ($\gamma^a$) was used by Dirac to describe the spin of fermions (spinors).  
ii. The second kind ($\tilde{\gamma}^a$) I am proposing to  explain the origin of  families~\footnote{ 
More about the two kinds of the Clifford algebra objects can be found in the 
refs.~\cite{norma,pikanorma,hn0203,NF}. The appendix~\ref{cliffordfamilies} gives a short 
overview of the technique.}. 
The theory predicts in the low energy regime two decoupled groups of four families. 
The lowest of the upper four families is the candidate to form the dark matter. 
Before the electroweak break there are four massless 
families of ($u^i$ and $d^i\,$, $i\in \{1,2,3,4\}$) quarks and 
($e^i$ and $\nu^i$) leptons,  left  handed weak charged and right handed weak chargeless, which 
when coupling to massive scalar fields with non zero vacuum expectation values and to gauge fields, 
become massive. 

Vector and scalar fields (both with respect to (3+1)), 
which originate in the spin connections of two kinds (they are gauge fields of the two kinds of 
the Clifford algebra objects)  and vielbeins at higher dimensions, are expected to have charges 
in the scalar, vector or any adjoint representations with respect to all charge groups, subgroups 
of the starting group. 
The question then arises: How can the scalar fields at low energies manifest effectively as 
weak doublets, as the Higgs certainly does? Although it is not at all simple to show, why the 
symmetries break in the way that they manifest the observed properties of fermions and vector and scalar 
bosons, the answer to the question, why scalar fields behave as weak doublets, is simple, 
if the way of breaking symmetries is assumed. 

In the refs.~\cite{pikanorma,NF,NBled2012} I present the assumed breaks of the starting 
symmetries of the simple 
starting action used by the {\it spin-charge-family} theory, as well as  possible 
answers  to the above questions.  
In this paper I repeat the assumptions of the {\it spin-charge-family} theory, 
I briefly represent the low energy manifestation of the simple starting action after the assumed 
breaks of the starting symmetries, analyse properties of the families: their charges, their coupling 
to the scalar fields,  and to the vector fields, their mass matrices and correspondingly their masses 
and mixing matrices  and masses of the weak bosons when they couple to several scalar fields 
(following to some extent the refs.~\cite{NF,NBled2012,NPLB,AN}),
I offer the answer to the question, why do scalar fields appear as doublets with 
respect to the weak $SU(2)$, while they behave as triplets with respect to the family groups, and I 
comment on predictions beyond the {\it standard model}.

Although the {\it  spin-charge-family} theory requires additional studies to be proved 
(or might be even disproved) that it is the right step beyond the {\it standard model}, 
the work done so far~\cite{norma,pikanorma,NF,gmdn,gn,GNBled2012,HN,NHD,AN} gives a real hope. 
The theory explains the assumptions of the {\it standard model}: 
The spins and charges of quarks and leptons, left and right handed, the families, the Higgs (manifesting 
as a superposition of several scalar fields), the Yukawa 
couplings, predicting the fourth family and the stable fifth family which forms the dark matter. It gives  
a hope, supported by the calculations done so far~\cite{GNBled2012,AN,gmdn,gn,NHD,HN}, that we shall 
understand the differences in properties of family members. 

The theory also predicts that since there are several scalar fields, which mass eigen states differ from 
the superposition with which they couple to different family members of different families and also to 
weak bosons,  future experiments will observe several scalar fields.

 %
 \section{Properties of fermions, gauge vector and scalar bosons}
 \label{SCFT}

 The starting symmetry $SO(13,1)$ breaks into $SO(7,1)\times$ $SU(3)\times U(1)_{II}$ 
 and then to $SO(3,1) \times SU(2)_{I}\times SU(2)_{II}$ $\times U(1)_{II}\times SU(3)$.
 In this paper we follow mainly the stage just before the electroweak break and after the 
 electroweak break. 
We shall call the break from $SU(2)_{I}\times SU(2)_{II}
\times U(1)_{II}$ to $SU(2)_{I}\times U(1)_{I}$ the  {\it break} $II$  and  the break from $SU(2)_{I} 
\times U(1)_{I}$ to $U(1)$, that is the electroweak break,  the  {\it break} $I$.

The  {\it spin-charge-family} theory~\cite{norma,pikanorma,NF,NBled2012,NPLB,GNBled2012,gmdn,gn,AN,HN,NHD,DNB,NHB}, 
a kind of the Kaluza-Klein-like theory but with families included, proposes in $d=(13+1)$ a simple action 
for a Weyl spinor and for the corresponding gauge fields  
\begin{eqnarray}
S            \,  = \int \; d^dx \;{\mathcal L}_{f} &+&  
 \int \; d^dx \; E\; (\alpha \,R + \tilde{\alpha} \, \tilde{R})\,,\nonumber\\
              {\mathcal L}_f = \frac{1}{2}\, (E\bar{\psi} \, \gamma^a p_{0a} \psi) &+&
              h.c.\,, 
\nonumber\\
p_{0a }        =  f^{\alpha}{}_a p_{0\alpha} + \frac{1}{2E}\, \{ p_{\alpha}, E f^{\alpha}{}_a\}_-\,,&&\quad   
   p_{0\alpha} =  p_{\alpha}  - 
                    \frac{1}{2}  S^{ab} \omega_{ab \alpha} - 
                    \frac{1}{2}  \tilde{S}^{ab}   \tilde{\omega}_{ab \alpha},                   
\nonumber\\ 
R              =   f^{\alpha [ a} f^{\beta b ]} \;(\omega_{a b \alpha, \beta} 
- \omega_{c a \alpha}\,\omega^{c}{}_{b \beta}) \} \,, && \quad 
\tilde{R}      =  f^{\alpha [ a} f^{\beta b ]} \;(\tilde{\omega}_{a b \alpha,\beta} - 
\tilde{\omega}_{c a \alpha} \tilde{\omega}^{c}{}_{b \beta})\,. 
\label{wholeaction}
\end{eqnarray}
The spin connection fields $\omega_{ab \alpha}$ are the gauge fields of the "charges" $S^{ab}=\frac{i}{4} 
(\gamma^a \gamma^b- \gamma^b \gamma^a)$, while $\tilde{\omega}_{a b \alpha}$ are the gauge fields of  the "family" 
quantum numbers $\tilde{S}^{ab}=\frac{i}{4} (\tilde{\gamma}^a \tilde{\gamma}^b- \tilde{\gamma}^b \tilde{\gamma}^a)$.

The fermion part of the action manifests after the breaks of symmetries to 
$SO(3,1)\times SU(2)_{I} \times SU(2)_{II}\times 
U(1)_{II}\times SU(3)$ as
\begin{eqnarray}
{\mathcal L}_f &=&  \bar{\psi}\gamma^{n} (p_{m}- \sum_{A,i}\; g^{A}\tau^{Ai} A^{Ai}_{m}) \psi 
+ \nonumber\\
               & &   \sum_{s=7,8}\,  \bar{\psi} \gamma^{s} \,(p_{s}- 
               \sum_{\tilde{A},i}\; \tilde{g}^{\tilde{A}} \tilde{\tau}^{\tilde{A}i} \tilde{A}^{\tilde{A}i}_{s} 
               - \sum_{A,i}\; g^{A}\tau^{Ai} A^{Ai}_{s}\;) \psi   + \nonumber\\
               & & {\rm the \;rest}\,, 
\label{faction}
\end{eqnarray}
where $n=0,1,2,3$ and
\begin{eqnarray}
\tau^{Ai} = \sum_{a,b} \;c^{Ai}{ }_{ab} \; S^{ab}\,, \quad  
\{\tau^{Ai}, \tau^{Bj}\}_- = i \delta^{AB} f^{Aijk} \tau^{Ak}\,,\nonumber\\
\tilde{\tau}^{Ai} = \sum_{a,b} \;\tilde{c}^{Ai}{ }_{ab} \; \tilde{S}^{ab}\,, \quad  
\{\tilde{\tau}^{Ai}, \tilde{\tau}^{Bj}\}_- = i \delta^{AB} f^{Aijk} \tilde{\tau}^{Ak}\,.
\label{tau}
\end{eqnarray}
$\tau^{Ai}$ determine all the charges of fermions, $A^{Ai}_{m}, \,m\in \{0,1,2,3\};$ (they are superposition of  
$f^{\alpha}_{a}\, \omega_{bc \alpha}$) the corresponding gauge vector fields and $A^{Ai}_{s}, \,s\in \{7,8\};$  
the gauge scalar fields (superposition of $f^{\sigma}_{s} \omega_{tz \sigma}$), while $
\tilde{\tau}^{Ai}$ determine all the family quantum numbers (the family charges) of 
fermions and $\tilde{A}^{Ai}_{s}$ (superposition of $f^{\sigma}_{a} \tilde{\omega}_{bc \sigma}$) 
the corresponding gauge scalar fields. The way of breaking 
symmetries is assumed  in a way that it leads to low energy observable phenomena.

At this step, that is before the {\it break} $II$,  the action describes eight massless fermions, 
interacting with the i. massless vector 
triplet ($SU(2)_{II}$) gauge fields, massless vector  triplet ($SU(2)_{I}$) weak gauge fields, 
massless vector singlet ($U(1)_{II}$) gauge fields, massless vector octet ($SU(3)$) gauge fields, 
and ii.  scalar fields of both kinds, that is of $S^{ab}$ and of $\tilde{S}^{ab}$ origin. 
The term vectors and scalars determines the relation to $d=(3+1)$. 

The scalar fields of any origin, appearing in the second line of Eq.~(\ref{faction}), which would 
manifest as  four vectors with respect to $s=5,6,7,8$, manifest as doublets for $s=7,8$.
Namely, if $s$ would be allowed in Eq.~(\ref{faction}) to run within all the indices of $SO(4)$, 
that is, if $s\in \{5,6,7,8\}$, the scalar fields would be in the vector representation of the group 
$SO(4)$: $(\frac{1}{2}, \frac{1}{2})$. The choice of $s\in \{7,8\}$ forces them to behave as doublets with 
respect to the $SU(2)_{I}$,  that is with respect to the weak  subgroup. Correspondingly all the 
scalar fields behave as  doublets with respect to the weak charge, while they are triplets with 
respect to the family quantum numbers, generated by $\tilde{S}^{ab}$.

There are scalar fields ($\vec{\tilde{A}}^{2}_{s}\,,$ $\vec{\tilde{A}}^{\tilde{N}_{R}}_{s}\,,$ $s=7,8$) 
which, after gaining nonzero vacuum expectation values in the {\it break} $II$, determine masses 
of the upper four families and of  the gauge bosons $A^{2\pm}_{m}= 
\frac{1}{\sqrt{2}}\,(A^{21}_{m}\mp i A^{22}_{m})\,,$
$A^{Y'}_{m} = -\sin \theta_2 \,A^{4}_{m} + \cos \theta_{2}\, A^{23}_{m}$, while the gauge vector fields 
$A^{Y}_{m} = \cos \theta_2 \,A^{4}_{m} + \sin \theta_{2} \,A^{23}_{m}$ and $\vec{A}^{1}_{m}$ and the 
lower four families remain  massless (see footnote~\footnote{   
( We have: $Y = \tau^4 + \tau^{23}$, $Y'=-\tan^2 \theta_{2} \tau^4 + \tau^{23}$, $\vec{\tau}^{2}=$
$\frac{1}{2}\,(S^{58}+ S^{67}, S^{57}- S^{68}, S^{56}+ S^{78})\,,$ $\vec{\tau}^{4}=$
$-\frac{1}{3}\,(S^{9\,10}+ S^{11\,12}+  S^{13\,14})\,,$ $\vec{\tilde{\tau}}^{2}=$
$\frac{1}{2}\,(\tilde{S}^{58}+ \tilde{S}^{67}, \tilde{S}^{57}- \tilde{S}^{68}, \tilde{S}^{56}+ 
\tilde{S}^{78})\,,$ $\vec{\tilde{N}}_{R} =$ $ \frac{1}{2}\,(\tilde{S}^{23}-i \tilde{S}^{01},
\tilde{S}^{31}-i \tilde{S}^{02}, \tilde{S}^{12}-i \tilde{S}^{03})$.} for definitions of quantum numbers).

In the electroweak  {\it break} $I$  the scalar fields ($\vec{\tilde{A}}^{1}_{s}\,,$ 
$\vec{\tilde{A}}^{\tilde{N}_{L}}_{s}\,$, $A^{Q}_{s}\,,$ $A^{Q'}_{s}\,$ and $A^{Y'}_{s}\,,$ $s=7,8$) 
gain nonzero vacuum expectation values, determining masses of the lower four families  and  of
the gauge bosons $A^{1\pm}_{m}= \frac{1}{\sqrt{2}}\,(A^{11}_{m}\mp i A^{12}_{m})\,,$
$A^{Q'}_{m} = -\sin \theta_1 \,A^{Y}_{m} + \cos \theta_{1}\, A^{13}_{m}$, while 
$A^{Q}_{m} = \cos \theta_1 \,A^{Y}_{m} + \sin \theta_{1}\, A^{13}_{m}$ remains massless (see 
footnote~\footnote{   
( We have: $Q = Y + \tau^{13}$, $Q'=-\tan^2 \theta_{1} Y + \tau^{13}$, $\vec{\tau}^{1}=$
$\frac{1}{2}\,(S^{58}- S^{67}, S^{57}+ S^{68}, S^{56}- S^{78})\,,$ $\vec{\tilde{\tau}}^{1}=$
$\frac{1}{2}\,(\tilde{S}^{58}- \tilde{S}^{67}, \tilde{S}^{57}+ \tilde{S}^{68}, \tilde{S}^{56}- 
\tilde{S}^{78})\,,$ $\vec{\tilde{N}}_{L} =$ $ \frac{1}{2}\,(\tilde{S}^{23}+i \tilde{S}^{01},
\tilde{S}^{31}+i \tilde{S}^{02}, \tilde{S}^{12}+i \tilde{S}^{03})$.} for definitions of quantum numbers).

Let us denote the scalar fields  contributing to masses of the lower four families (scalar fields,
contributing to the {\it break} II are presented in the appendix~\ref{breakII} in 
Eq.~(\ref{allscalarsII})) and to the 
weak bosons masses with a common vector
%
\begin{eqnarray}
\label{allscalars}
\Phi^{I\,Ai} & \equiv&  \Phi^{I\, Ai}_{\mp}\,, \quad  
\Phi^{I\, Ai}_{\mp}= (\vec{\tilde{A}}^{1}_{\mp}\,,\vec{\tilde{A}}^{\tilde{N}_L}_{\mp}\,,
A^{Y'}_{\mp}\,,A^{Q'}_{\mp}\,,A^{Q}_{\mp})\,, \nonumber\\
\Phi^{Ai}_{\mp}= (\Phi^{Ai}_{7} \pm i \Phi^{Ai}_{8})\,, &&\quad A_{I}= \{\tilde{1}, \tilde{N}_L,Y',Q',Q\}\,.
\end{eqnarray}
We  choose a renormalizable effective potential $V(\Phi^{I,Ai})  $ for the (assumed to be) real 
scalar fields $\Phi^{I,Ai}$ (Eq.~(\ref{allscalars})), which couple among themselves 
%
%
\begin{eqnarray}
\label{veff}
V(\Phi^{I\,Ai})   &=& \sum_{A,i}\{\, -\frac{1}{2} \,  (m^{I}_{Ai})^2 (\Phi^{I\,Ai})^2 + 
\frac{1}{4}\, \sum_{ B,
j}\, \lambda^{I\,Ai\,Bj}\, (\Phi^{I\,Ai})^2 \, (\Phi^{I\,Bj})^2\}\,.
\end{eqnarray}
Couplings among the scalar fields are here chosen to be symmetric: $\lambda^{Ai\,Bj}=\lambda^{Bj\, Ai}$.

Scalar fields couple to the gauge bosons  at the {\it break } $I$ 
according to the Lagrange function  ${\mathcal L}_{s\,I}$
%
\begin{eqnarray}
\label{scalarLagrange0I}
{\mathcal L}_{s\,I} &=& \sum_{A,i}\,\,(p_{0m} \Phi^{I\,Ai}_{})^{\dagger}(p_{0}{}^{m}\, \Phi^{I \,Ai}_{})
- V(\Phi^{I\,Ai})\,,
\nonumber\\
p_{0m} &=& p_{m} - \{ g^{Y}\ \,\tau^{Y}\,A^{Y}_{m} +  g^{1}\, \vec{\tau}^1\, \vec{A}^{1}_{m} \}\,. 
\end{eqnarray}
We shall study properties of scalar fields in subsection~\ref{propertiesscalars} and 
section~\ref{minimization}.

Expressing the operators $\gamma^7$ and $\gamma^8$ in terms of the nilpotents $\stackrel{78}{(\pm)}$ 
(appendix~\ref{cliffordfamilies}), the second line of Eq.~(\ref{faction}) can be rewritten as follows
\begin{eqnarray}
\bar{\psi} M \psi   &=& \sum_{s=7,8}\,\bar{\psi} \gamma^{s}\, p_{0s}\,\psi  =
 \psi^{\dagger}\, \gamma^{0}\, (\stackrel{78}{(-)}\,p_{0-} + \stackrel{78}{(+)}\,p_{0+}) \psi\, ,\nonumber\\ 
\stackrel{78}{(\pm)}&=&  \frac{1}{2}\,(\gamma^{7} \, \pm i\,\gamma^{8} )\,, \nonumber\\ 
p_{0\pm}&=& (p_{07} \mp i\, p_{08})\,, \quad p_{0\pm}= p_{\pm} - \frac{1}{2}\, S^{ab} \omega_{ab \pm} -
\frac{1}{2}\, \tilde{S}^{ab} \tilde{\omega}_{ab \pm}\,\nonumber\\
\omega_{ab \pm}&=& (f^{\sigma}_{7} \mp i\, f^{\sigma}_{8})\,\omega_{ab \sigma}\,, \quad
\tilde{\omega}_{ab \pm}= (f^{\sigma}_{7} \mp i f^{\sigma}_{8})\,\tilde{\omega}_{ab \sigma}\,. 
\label{factionM}
\end{eqnarray}
In the mass term $\bar{\psi} M \psi  $ there is $\stackrel{78}{(-)}$ which transforms a weak chargeless 
$u_{R}^i$ ($\nu_{R}^i$)  with $Y=\frac{2}{3}$ ($Y=-\frac{1}{2}$) into the corresponding left handed 
members and it is $\stackrel{78}{(+)}$ which transforms a weak chargeless $d_{R}^i$ ($e_{R}^i$)  
with $Y=-\frac{1}{3}$ ($Y= -\frac{1}{2}$) into the corresponding left handed  members for each family
separately. And there are $\vec{\tilde{\tau}}^{A}$ which transform a family member of one family to 
the same family member of another family. 


There are eight massless families before the {\it break} $II$. 
Four families out of eight are doublets with respect to 
$\vec{\tilde{\tau}}^{2}$ (one $\tilde{SU}(2)$ subgroup of $\tilde{SO}(4)$, see footnotes for the definition) 
and $\vec{\tilde{N}}_{R}$ (one  $\tilde{SU}(2)$ subgroup of $\tilde{SO}(3,1)$, see footnotes for the 
definition). 
The rest four families are doublets with respect to (the rest $\tilde{SU}(2)$ subgroup of $\tilde{SO}(4)$) 
$\vec{\tilde{\tau}}^{1}$ and (the rest  $\tilde{SU}(2)$ subgroup of $\tilde{SO}(3,1)$) $\vec{\tilde{N}}_{L}$.

The first group of four families become massive,  when the gauge scalar fields (the superposition of 
the  scalar fields $\tilde{\omega}_{ab s}$) of the charges $\vec{\tilde{\tau}}^{2}$ and $\vec{\tilde{N}}_{R}$  
gain nonzero vacuum expectation values. 
Since family quantum numbers do not distinguish among family members, the masses of $u^{II\,i}$, 
$d^{II\,i}$, $\nu^{II\,i}$ and $e^{II\,i}$ are  the same on the tree level, if 
$\Phi^{II\,Ai}_{+}=$ $\Phi^{II\,Ai}_{-}$.

The second group of four families become massive, when  gauge scalar fields (superposition of  
$\tilde{\omega}_{ab s}$) of the charges $\vec{\tilde{\tau}}^{1}$ and $\vec{\tilde{N}}_{L}$, 
together with those (superposition of  $\omega_{sts'}$)  of the $U(1)$ charges ($Y',Q', Q$)  
gain nonzero vacuum expectation values.

The mass term $\bar{\psi}\,  M \, \psi $  determines in this case the tree level mass matrices 
of quarks and leptons of the lower four families, among which are the measured three families, 
and correspondingly the masses, the Yukawa couplings and the mixing matrices for these four 
families after the loop corrections are taken into account.  The tree level mass term 
distinguishes among the members of any of the four families: Due to the eigen values of  
the operators $Q\,,Q'$ and $Y'$,  and due to $\stackrel{78}{(\mp )}\Phi^{I\, Ai} $, since 
(by assumption) $\Phi^{I\,Ai}_{+}\ne$ $\Phi^{I\,Ai}_{-}$. 
The fields  $\tilde{A}^{\tilde{N}_L i}_{\mp}$ and  $\tilde{A}^{1i}_{\mp}$ are  
triplets with respect to the two $\tilde{SU}(2)$  family subgroups ($\vec{\tilde{\tau}}^{1}$,
$\vec{\tilde{N}}_{L}$), while they are doublets with respect to the weak charge (see 
subsection~\ref{propertiesscalars}).

After the electroweak break the effective Lagrange density for the lower four families of fermions  
looks~\cite{NF}  as~(Eqs.~(\ref{faction}, \ref{factionM}))
 \begin{eqnarray}
 {\mathcal L}_{I\,f} &=&  \bar{\psi}\, (\gamma^{m} \, p_{0m} - \, (\stackrel{78}{(-)}\,p_{0-}
           + \stackrel{78}{(+)}\,p_{0+})) \psi\, , \nonumber\\
          p_{0m}&=& p_{m} - \{ g^{Q}\,Q\,A_{m} +  g^{Q'}\,Q'\, Z^{Q'}_{m} +
          \frac{g^{1}}{\sqrt{2}}\, (\tau^{1+} \,W^{1+}_{m} + \tau^{1-} \,W^{1-}_{m}) + 
          g^{Y'} \,Y'\, A^{Y'}_{m}          \,\} \nonumber\\
          p_{0\pm}&=& p_{\pm} - \{ \tilde{g}^{\tilde{N}_L} \,\vec{\tilde{N}}_L\, 
          \vec{\tilde{A}}^{\tilde{N}_L}_{\pm} +
 		  	                \tilde{g}^{\tilde{Q}'} \, \tilde{Q}'\,
 		  	                \tilde{A}^{\tilde{Q}'}_{\pm}
 		  	       	          + \frac{\tilde{g}^{1}}{\sqrt{2}}\, (\tilde{\tau}^{1+}
 		  	       	          \,\tilde{A}^{1+}_{\pm}
 	          + \tilde{\tau}^{1-}\,  \tilde{A}^{1-}_{\pm})\nonumber\\ 	
 	          &+&  g^{Q} 
 	          \, Q\, A^{Q}_{\pm} +  g^{Q'}_{1}\, 
 	          Z^{Q'}_{\pm} +
               g^{Y'}_{1} Y'\, A^{Y'}_{\pm}\,\}\,,\nonumber\\ 
 	           Q&=& \tau^{13} + Y\,,\quad Q'_{1}= (\tau^{13} - \tan^2 \theta_1 \,Y)\,, 
 	           \quad Y'_{1}= (\tau^{23} - \tan^2 \theta_2 \,\tau^4)\,.
 \label{factionI}
 \end{eqnarray}
$Q$ and $Q'$  are the {\it standard model} like  charges ($ \vartheta_{1}$ does not need to be $\theta_{1}$), 
$Y'$ is the additional quantum number~\cite{NF},  appearing in the {\it spin-charge-family} theory 
 (similarly as it does in the $SO(10)$ models), and $g^{Q}= g^{Y}\,\cos \theta_1$, 
 $g^{Q'}= g^{1}\,\cos \theta_1$ and  $g^{Y'}= g^{4}\,\cos \theta_2$.

\subsection{Transformation properties of scalar fields}

\label{propertiesscalars}

In the {\it standard model} the Higgs and the anti-Higgs (a kind of Majorana)  "dress" 
the right handed fermions with the appropriate weak and hypercharge so that the "dressed" ones 
carry quantum numbers of the left handed partners.

How can these properties of the Higgs be explained in the {\it spin-charge-family} theory?
It is the second summand in Eq.~(\ref{faction}) which determines on the tree level, after  
scalar fields  gain non zero vacuum expectation values,  masses of fermions.  
In the case that all four indices $s=(5,6,7,8)$ would be allowed, all the scalar fields would 
transform as vectors (as well known from the vector representations in $d=(3+1)$) according to the 
($\frac{1}{2},\frac{1}{2}$) representation. Since only $s=(7,8)$ is allowed, they transform 
as doublets with respect to the weak $SU(2)$. 

We also notice that the operators $\gamma^{0}\, \stackrel{78}{(\mp)}\,\tau^{Ai}\,\Phi^{Ai}_{\mp} $ 
transform right handed weak chargeless fermions of a particular hyper charge into the corresponding 
left handed weak charged partners
\begin{eqnarray}
\label{howStransform}
\gamma^0 \stackrel{78}{(-)} \tau^{Ai} \Phi^{Ai}_{-} \psi_{(u, \nu)R} 
&&\rightarrow \tau^{Ai} \Phi^{Ai}_{-} \psi_{(u, \nu)L}\,, \nonumber\\
\gamma^0 \stackrel{78}{(+)} \tau^{Ai} \Phi^{Ai}_{+} \psi_{(d, e)R} 
&&\rightarrow \tau^{Ai} \Phi^{Ai}_{+} \psi_{(d, e)L}\,.
\end{eqnarray}
Here $\tau^{Ai}$ stay for all the quantum numbers of either $\tilde{S}^{ab}$ 
( in the case of the lower four families these are $\vec{\tilde{\tau}}^{1}$ and $\vec{\tilde{N}}_{L}$,  
causing transitions among families) 
or $S^{ab}$ ($Y',Q'$ and $Q$, which are diagonal: In the family member and the family quantum numbers), 
while $ \Phi^{Ai}_{\mp}$ (Eq.~(\ref{allscalars})) stay for all the corresponding scalar gauge fields of either 
$\tilde{\omega}_{st\mp}$ or $\omega_{st\mp}$ origin.

Eq.~(\ref{howStransform}) demonstrates that all the scalar fields transform as doublets with 
respect to the weak charge, each of the two kinds ($ \Phi^{Ai}_{-}$,  $ \Phi^{Ai}_{+}$) having 
different hyper charges ($\Phi^{Ai}_{-}$ is a member of a doublet with the 
weak charge $\frac{1}{2}$ and the hypercharge $-\frac{1}{2}$\,, while $\Phi^{Ai}_{+}$ is  a 
member of a doublet with the weak charge $-\frac{1}{2}$ and the hypercharge $\frac{1}{2}$), 
the same for all the families. 
Their nonzero vacuum expectation values do not need to be the same: ($< \Phi^{Ai}_{-}> \ne $ 
$< \Phi^{Ai}_{+}>$, for any $(A,i)$). In this case the pair with nonzero vacuum expectation values 
does not look like a Majorana.

Scalar fields which change properties of the upper four families do not distinguish on the tree level 
among family members provided that $<\vec{\tilde{A}}^{2}_{-}>= 
<\vec{\tilde{A}}^{2}_{+}>$ and $<\vec{\tilde{A}}^{\tilde{N}_R}_{-}>= 
<\vec{\tilde{A}}^{\tilde{N}_R}_{+}>$ for each vector component. In this case all the family members of a 
particular family have on the tree level the same mass.  
They keep correspondingly the $SU(2)_{I}$ (the weak charge) symmetry unchanged, although 
the right handed partners (having the same mass as the left handed ones) do not carry the weak charge.  

The scalar fields  (
$\vec{\tilde{A}}^{1}_{\pm}\,,\vec{\tilde{A}}^{\tilde{N}_{L}}_{\pm}$, all transforming  as doublets 
with respect to the weak charge group) transform as triplets with respect to the $\tilde{SU}(2)$ groups 
(the generators of which are $\vec{\tilde{\tau}}^{1}\,,$ $\vec{\tilde{N}}_{L} $), while they transform 
as singlets with respect to $U(1)$ groups (the generators of which are $Y' \,,Q'\,,$and $ Q$).

Scalar fields with nonzero vacuum expectation values obviously change properties of the vacuum.
%
There are several changes of the vacuum during several breaks of symmetries of the starting one: 
$SO(13,1)$. Let us concentrate on {\it break } $I$. The {\it break} II is discussed 
in appendix~\ref{breakII}.

At the electroweak {\it break } $I$ 
the vacuum  changes again. 
The mass term~(Eq.~(\ref{factionM})) forces all the scalar fields, which obtain nonzero vacuum expectation 
values at this break,  to be doublets with respect to the $SU(2)_{I}$ (weak) charge, while they keep 
their adjoint representations with respect to all the charges of the $\tilde{S}^{ab}$ origin 
($\vec{\tilde{\tau^1}}$, $\vec{\tilde{N}}_{L}$). 
To the vacuum 
two new kinds of terms must be added
\begin{eqnarray}
\label{vacuumI}
\stackrel{78}{(-)} \ominus_{I}&=&
\stackrel{78}{(-)}
Ts_{\tilde{N}_L} 
 | 
(\stackrel{56}{[+]} \stackrel{78}{(+)})
\,Td_{(-)\tilde{\tau}^1}
||\stackrel{9\,10}{[+]}\stackrel{11\,12}{[+]} \stackrel{13\,14}{[+]}\,,
\nonumber\\
\stackrel{78}{(+)} \oplus_{I}&=&
\stackrel{78}{(+)}
Ts_{\tilde{N}_L} 
 | 
(\stackrel{56}{[-]} \stackrel{78}{(-)})
\,Td_{(+) \tilde{\tau}^1}
||\stackrel{9\,10}{[-]}\stackrel{11\,12}{[-]} \stackrel{13\,14}{[-]}
\,.
\end{eqnarray}
Here $Ts_{\tilde{N}_L} $ denotes a triplet with respect to the operators $\vec{\tilde{N}}_{L}$ and 
a singlet with respect to $\vec{N}_{L}$, while $(\stackrel{56}{[+]} \stackrel{78}{(+)})
\,Td_{(\mp)\tilde{\tau}^1}$ are the two triplets with respect to $\vec{\tilde{\tau}}^{1}$ and  doublets with respect 
to $\vec{\tau}^1$. 

One finds
\begin{eqnarray}
\label{propvacuumI}
\tau^{1+} \tau^{1-}\,\stackrel{78}{(+)} \oplus_{I}  = \stackrel{78}{(+)} \oplus_{I}\,,&&
\tau^{1-} \tau^{1+}\,\stackrel{78}{(-)} \ominus_{I} = \stackrel{78}{(-)} \ominus_{I}\,,\nonumber\\
Q \,\stackrel{78}{(+)} \oplus_{I}  =  &0& = Q \,\stackrel{78}{(-)} \ominus_{I}\,, \nonumber\\
Q' \,\stackrel{78}{(+)} \oplus_{I} &=& -\frac{1}{2\cos^2 \theta_1}\,,\quad 
Q' \,\stackrel{78}{(-)} \ominus_{I} = \frac{1}{2\cos^2 \theta_1}
\,.
\end{eqnarray}
We see that in the {\it break} $I$ the fields $A^{1\pm}_m(= W^{\pm}_{m})$ 
and $A^{Q'}_{m} \,(=Z_m) = \cos \theta_1 A^{13}_m - \sin \theta_1 A^{Y}_{m}$ become massive, while 
$A^{Q}_{m}\,(= A_{m})= \sin \theta_2 A^{13}_m + \cos \theta_1 A^{Y}_{m}$  stays massless, 
provided that $\frac{g^1}{g^Y} \,\tan \theta_1=1$. 

In the ref.~\cite{gn} properties of the stable fifth family  members were studied and the possibility 
that they form a dark matter analysed.
We study in this paper properties of scalar and vector gauge fields and of families of quarks and 
leptons  for (mostly) the lower four families.

\section{Scalar fields in minimization procedure on tree level}
\label{minimization}

Let us look for the minimum of the  two potentials presented in Eq.~(\ref{veff}) and search for the 
mass eigen states 
on the tree level~\cite{NPLB}.  Let $\Phi^{Ai}$ denotes  $\Phi^{I\,Ai}$ (or $\Phi^{II\,Ai}$) and $V(\Phi^{Ai})$
the corresponding effective potential. 
First we look for the first derivatives with respect to all the interacting scalar fields and put them equal 
to zero
\begin{eqnarray}
\label{scalarminimum}
\frac{\partial V(\Phi^{Ai})}{\partial \Phi^{Ai}} &=& 0 = \Phi^{Ai}\,[- (m_{Ai})^2 + 
\lambda^{Ai} (\Phi^{Ai})^2  + 
\sum_{B, j}\, \lambda^{Ai\,Bj} \,(\Phi^{Bj})^2]\,.
\end{eqnarray}
Here the notation $\lambda^{Ai}: = \lambda^{AiAi}$ is used. When expressing the minimal 
values of the scalar fields, let as call them $v_{Ai}$, 
as functions of the parameters, Eq.~(\ref{scalarminimum}) leads to the  coupled 
equations for the same number of  unknowns $v_{Ai}= \Phi^{Ai}_{min}$  
\begin{eqnarray}
\label{scalarv0}
- (m_{Ai})^2 + \sum_{B,j}\,\lambda^{Ai Bj} (v_{Bj})^2=0\,.
\end{eqnarray}
Looking for the second derivatives at the minimum determined by $v_{Ai}$ one finds
\begin{eqnarray}
\label{scalarsecderv0}
\frac{\partial^2 V(\Phi^{Ck})}{\partial \Phi^{Ai}\partial\Phi^{Bj} }|_{ v_{Ck}} &=& 
2 \lambda^{Ai Bj} v_{Ai}\,v_{Bj}\,. 
\end{eqnarray}
Let us  look for the basis  $\Phi^{\rho}$ (we should keep in mind  the index ($\pm$), 
although we do not write it down)  
\begin{eqnarray}
\label{scalarnew}
\Phi^{Ai} = \sum_{\rho}\, {\cal C}^{Ai}_{\rho}\, \Phi^{\rho}\,,
\end{eqnarray}
in which on the tree level the potential would be diagonal 
\begin{eqnarray}
\label{scalardiag}
V(\Phi^{\rho})   &=& \sum_{\rho}\{\, - \frac{1}{2} \, (m_{\rho})^2 (\Phi^{\rho})^2 + 
\frac{1}{4}\, \lambda^{\rho}\, (\Phi^{\rho})^4\}\,,
\end{eqnarray}
with $\frac{\partial V}{\partial \Phi^{\rho}}|_{v_{Ai}} =$ $  
  \sum_{Ai} \, \frac{\partial V}{\partial \Phi^{Ai}}
 \frac{\partial \Phi^{Ai}}{\partial \Phi^{\rho}}|_{v_{Ai}} = 0\,,$ 
 with $\Phi^{\rho}_{min}$ at these points called 
$v_{\rho}$ $=\sum_{Ai}\, {\cal C}^{Ai\, T}_{\rho}\, v_{Ai}$ ($T$ denotes transposition), 
and correspondingly with  $\frac{\partial^2 V}{\partial (\Phi^{\rho})^2}|_{ v_{\rho}}$
= $- (m_{\rho})^2 + 3 \lambda^{\rho }\, (\Phi^{\rho})^2|_{ v_{\rho}}
$= $2 \,\lambda^{\rho} \,(v_{\rho})^2 =$ $\sum_{A,i,B,k}\, 2 \lambda^{Ai Bj} \,v_{Ai}\,v_{Bj}\,
{\cal C}^{Ai}_{\rho}\,{\cal C}^{Bj}_{\rho}$.
This means that the new basis can be found by diagonalizing the matrix of the second derivatives 
at the minimum and correspondingly put to zero the determinant
\begin{eqnarray}
\label{scalardiag1}
\det \left(\begin{array}{cccc}
2 \lambda^{\tilde{N}_{L} 1}\, \, (v_{\tilde{N}_{L} 1})^2 - 2\,(m_{\rho})^2,&
2 \lambda^{\tilde{N}_{L} 1 \, \tilde{N}_{L} 2}\,v_{\tilde{N}_{L} 1}\, v_{\tilde{N}_{L} 2},&
2 \lambda^{\tilde{N}_{L} 1 \, \tilde{N}_{L} 3}\,v_{\tilde{N}_{L} 1}\, v_{\tilde{N}_{L} 3},&\ldots\\
2 \lambda^{\tilde{N}_{L} 2 \, \tilde{N}_{L} 1}\,v_{\tilde{N}_{L} 2}\, v_{\tilde{N}_{L} 1},&
2 \lambda^{\tilde{N}_{L} 2}\, \, (v_{\tilde{N}_{L} 2})^2-2\, (m_{\rho})^2,&
2 \lambda^{\tilde{N}_{L} 2 \, \tilde{N}_{L} 3}\,v_{\tilde{N}_{L} 1}\, v_{\tilde{N}_{L} 3}, &\ldots\\
\vdots&\vdots&\ddots
\end{array}\right)\,.
\end{eqnarray}
The same number of orthogonal scalar fields $\Phi^{\beta}$, with nonzero 
vacuum expectation values and nonzero masses, as we started with, follow. 
To each of them one eigen value $2 (m_{\rho})^2$ corresponds, determined by the parameters
$m_{Ai}$ and $\lambda^{AiBj}$ of Eq.~(\ref{veff}).

For the time evolution of the free scalar fields one correspondingly finds for each $\beta$
\begin{eqnarray}
\label{scalaralphatimeevol}
\Phi^{\beta}(t) =  e^{-i m_{\beta} (t-t_0)}\, \Phi^{\beta}(t_0)\,.  
\end{eqnarray}
\subsection{A simple example}
\label{example2x2}

Let us examine a simple case,  one triplet, say $\vec{\tilde{A}}^{1}$, and let us call these 
three scalar states 
$\Phi^{i}$. 
Following Eq.~(\ref{scalarminimum}) one obtains 
$- (m_{i})^2 + \sum_{j}\,\lambda^{i j} (v_{j})^2=0\,,\quad {\rm for\;\;each\;\;} \,i=1,2,3$.  
Let us further simplify the example by the assumption that one of these three fields is decoupled: 
$\lambda^{i3}=0$, for $i=(1,2)$. 
Then it follows  for the vacuum expectation values $v_{i}, i\in \{1,2,3\}$
\begin{eqnarray}
\label{scalarv02}
(v_{1})^2 &=& \frac{- \lambda^{12} (m_2)^2 + \lambda^{^2} (m_1)^2}{\lambda^1 \lambda^2- (\lambda^{12})^2}\,,\quad
(v_{2})^2  = \frac{- \lambda^{12} (m_1)^2 + \lambda^{^1} (m_2)^2}{\lambda^1 \lambda^2- (\lambda^{12})^2}\,,\quad
(v_{3})^2  = \frac {(m_3)^2}{\lambda^3}\,.
\end{eqnarray}
The second derivatives at the minimum, 
$\frac{\partial^2 V(\Phi^{k})}{\partial \Phi^{i}\partial\Phi^{j} }|_{\Phi^{k}= v_{k}} = 
2 \lambda^{i j} v_{i}\,v_{j}\,$,  
lead to the determinant~(Eq.(\ref{scalardiag1})), 
from where one obtains the eigen masses
\begin{eqnarray}
\label{scalardiagres}
(m^{1,2})^{2}&=& \frac{1}{2}\,\{[\lambda^1 (v_1)^2 + \lambda^2 (v_2)^2] \mp 
\sqrt{[\lambda^2 (v_2)^2 - \lambda^1 (v_1)^2]^2+4 (\lambda^{12})^2 \,(v_1)^2\, (v_2)^2}\}\,,
\end{eqnarray}
and $(m^3)^2 = (m_{3})^2$. 
If the coupling between the two scalar components is zero, the trivial case of three uncoupled  
scalar fields  follows. In the case that the two masses, $m_{1}$ and  $m_{2}$,  are equal and that also 
the two self strengths are the same, $\lambda^1=\lambda^2$, then $(v_1)^2 = (v_2)^2 $  and the 
two eigen values for masses are  $(m^{1,2})^{2}=(v_1)^2 [(\lambda^1- \lambda^{12}), (\lambda^1+ 
\lambda^{12}] $. In the case that $\lambda^1$ and $\lambda^{12}$ are close to each other, 
the two eigen values differ a lot. 
In the case of $\lambda^{12}=0$ the two scalars would manifest and be observed as only 
one. 

Such a simplified situation illustrates that the mass eigen states  of the scalar fields might 
differ a lot from the superposition of the scalar fields which couples to any of the family members of any 
of the families, the tree level mass matrices of which are presented in 
Table~\ref{Table VIII.} and in Eq.~(\ref{adiag}) and discussed in next section~\ref{fermionmassmatrix}.

\section{Coupling of family members to  scalar fields}
\label{fermionmassmatrix}

There is only explanation for the dark matter~\cite{gn} as the clusters of the stable fifth family 
members which supports so far the existence of the upper four families. 
To differences in masses of the family members of the lower four families 
contribute on the tree level not only the differences in the expectation values  $<\Phi^{I\, Ai}_{-}> \ne$
$ <\Phi^{I\, Ai}_{+}>$,  $\Phi^{I\, Ai}_{\mp}=$  ($\vec{\tilde{A}}^{1}_{\mp},$  
$ \vec{\tilde{A}}^{\tilde{N}_{L}}_{\mp}$), but also  the scalar fields ($A^{Y'}_{\mp}$, $A^{Q'}_{\mp}$,
$A^{Q}_{\mp}$), which couple to each member of a family in a different way.

The tree level contributions of $\vec{\tilde{A}}^{\tilde{N}_{L}}_{\mp}$  and 
$\vec{\tilde{A}}^{1}_{\mp}$ (Eq.~(\ref{factionI})) to the mass matrix of any family member 
look~\cite{NF,AN} as it is presented in Table~\ref{Table VIII.}. The notation 
$\tilde{a}^{\tilde{A}i}_{\mp}=$ $-\tilde{g}^{\tilde{A}}\, 
\tilde{v}_{\tilde{A}i\,\mp}$ is used, where $\tilde{v}_{\tilde{A}i\,\mp}$ are the vacuum expectation 
values of the corresponding scalars ($<\tilde{A}^{Ai}_{\mp}>$). Let us repeat that $\tilde{a}^{\tilde{A}i}_{\mp}$ 
distinguish among ($u^{i}\,, \nu^{i}$) ($(-)$) and ($d^{i}\,, e^{i}$) ($(+)$).

The contributions of  $ g^{Q}_{1}\,,    
Q\,A^{Q}_{\mp}\,, $ $g^{Q'}_{1}         
Q'\, Z^{Q'}_{\mp}\,,$ and $g^{Y'}\,Y'\, A^{Y'}_{\mp}$ are not presented in Table~\ref{Table VIII.}. 
They  are  different  for each of the family member $\alpha =(u^i, d^i, \nu^i, e^i)$ and the same for 
all the families ($i = (1,2,3,4)$) 
\begin{eqnarray}
\label{adiag}
a^{\alpha}_{\mp}&=&-\{ g^{Q}_{1}\,Q^{\alpha}\,v_{Q \,\mp} + g^{Q'}_{1}\,Q^{'\alpha}\,v_{Q'\,\mp}
+ g^{Y'}\,Y^{'\alpha}\,v_{Y'\,\mp} 
\}\,, 
\end{eqnarray}
with $Q^{\alpha}\,,$ $Q^{'\alpha}$ and $Y^{'\alpha}$, 
which  are eigen values of the corresponding operators for the family member state $\alpha$.

While the mass matrix elements might (and are expected to) change considerably when 
higher order corrections are taken into account, we expect 
that the symmetries stay  the ones of the tree level contributions, the same for all the family members,
as presented on Table~\ref{Table VIII.}. 
 \begin{table}
 \begin{center}
\begin{tabular}{|r||c|c|c|c||}
\hline
 $i$&$ 1 $&$ 2 $&$ 3 $&$4 $\\
\hline\hline
$1 $&
$ - \frac{1}{2}\,( \tilde{a}^{13}_{\mp} + \tilde{a}^{\tilde{N}^{3}_{L}}_{\mp})$&
$\tilde{a}^{\tilde{N}_{L}^{-}}_{\mp}$&$0$&
$   \tilde{a}^{1-}_{\mp}$  \\
\hline
$2$ &  $ \tilde{a}^{\tilde{N}_{L}^{+}}_{\mp} $ &
$ \frac{1}{2}( -\tilde{a}^{13}_{\mp } + \tilde{a}^{\tilde{N}^{3}_{L}}_{\mp}) $&
$\tilde{a}^{1-}_{\mp}   $ &$0$\\
\hline
$3$ & $0$& $\tilde{a}^{1+}_{\mp}$&
$  \frac{1}{2}\,( \tilde{a}^{13}_{\mp}  - \tilde{a}^{\tilde{N}^{3}_{L}}_{\mp}) $ &
 $\tilde{a}^{\tilde{N}_{L}^{-}}_{\mp}$ \\
\hline
$4$ & $\tilde{a}^{1+}_{\mp}$& $0$&$  \tilde{a}^{\tilde{N}_{L}^{+}}_{\mp}  $ &
 $\frac{1}{2}\,( \tilde{a}^{13}_{\mp}  + \tilde{a}^{\tilde{N}^{3}_{L}}_{\mp}) $
\\
\hline\hline
\end{tabular}
 \end{center}
 \caption{\label{Table VIII.}  The  contributions of the fields ($-\tilde{g}^{1}\, 
 \vec{\tilde{\tau}}^{1}\,\vec{\tilde{A}}^{1}_{\mp}\,,
 \, -\tilde{g}^{\tilde{N}_{L}}\, \vec{\tilde{N}}^{i}_{L}\, \vec{\tilde{A}}^{\tilde{N}_L}_{\mp}$) 
 to the mass matrices on the tree level (${\cal M}_{(o)}$) for the lower four  families  of quarks and
 leptons after the electroweak break are presented.  
The notation $\tilde{a}^{\tilde{A}i}_{\mp}=$ $-\tilde{g}^{\tilde{A}}\, \tilde{v}^{\tilde{A}i}_{\mp}$ is used. 
 }
\end{table}
Loop corrections, to which also the massive gauge fields and dynamical massive scalar fields contribute,  
are expected to strongly  influence fermions properties. 
These calculations are in progress~\cite{AN,GNBled2012} and look so far very 
promising in offering the right answers for the masses and mixing matrices of all the family members:
quarks and leptons, treating colour charge and colour chargeless members in an equivalent way. 
It turns out that both, quarks and leptons, mass matrices 
behave in a very similar way. No additional neutrinos, offering a "sea-saw" mechanism, are needed.

Let $\psi^{\alpha}_{(L,R)}$ denote massless and $\Psi^{\alpha}_{(L,R)}$  massive four vectors 
for each family member  $\alpha= (u_{L,R}, d_{L,R}, \nu_{L,R}, e_{L,R})$ after taking into account 
loop corrections in all orders~\cite{NF,AN},  
$\psi^{\alpha}_{(L,R)} = V^{\alpha}_{(L,R)} \,\Psi^{\alpha}_{(L,R)} \,$,
and let
$(\psi^{\alpha \,k}_{(L,R)}\,$,  $\,\Psi^{\alpha\, k}_{(L,R)} )$ 
be any component of the four vectors, massless and massive, respectively.
On the tree level  we have 
$\psi^{\alpha}_{(L,R)}=V^{\alpha}_{(o)}\:
\Psi^{\alpha \,(o)}_{(L,R)}$ 
and
\begin{equation}
\label{treenotation}
          < \psi^{\alpha}_{L}|\gamma^0 \, {\cal M}^{\alpha}_{(o)}\,
 |\psi^{\alpha}_{R}> = < \Psi^{\alpha \,(o)}_{L}|\gamma^0 \,V^{\alpha\,
 \dagger}_{(o)}\, {\cal M}^{\alpha}_{(o)}\,V^{\alpha}_{(o)}\,|\Psi^{\alpha}_{R \,(o)}>,
\end{equation}
with 
${\cal M}^{\alpha}_{(o)k\, k'}=\sum_{A,i}\, (-g^{Ai} \, v_{Ai\, \mp})\,\, C^{\alpha}_{k\,k'}\,$. 
The coefficients $ C^{\alpha}_{k\,k'}$ can be read from  Table~\ref{Table VIII.}. 
It then follows
\begin{eqnarray}
\label{Phipsi}
\overline{\Psi}^{\alpha}\,V^{\alpha \dagger}_{(o)}\, {\cal M}^{\alpha}_{(o)}\,V^{\alpha }_{(o)}\:
 \Psi^{\alpha} &=& \overline{\Psi}^{\alpha}\,{\rm diag}(m^{\alpha}_{(o)1}\,,\cdots\,,m^{\alpha}_{(o)4})\,
 \Psi^{\alpha}\,, \nonumber\\
V^{\alpha \dagger}_{(o)}\, {\cal M}^{\alpha}_{(o)}\,V^{\alpha }_{(o)}&=& \Phi^{\alpha}_{f(o)}\,.
\end{eqnarray}
The coupling constants $m^{\alpha}_{(o)k}$ (in some units) of the dynamical scalar fields 
$\Phi^{\alpha}_{f(o) \,k}$ to 
the family member  $\Psi^{\alpha \,k}$ belonging to the $k^{th}$ family are on the tree level 
correspondingly equal to
\begin{eqnarray}
\label{Phipsiex}
 (\Phi^{\alpha}_{\Psi(o)})_{k\,k'}\,\Psi^{\alpha\,k'} &=& \delta_{k\,k'}
 \,m^{\alpha}_{(o)k}\,\Psi^{\alpha\,k}\,. 
\end{eqnarray}

The superposition of scalar fields $(\Phi^{\alpha}_{f(o)})$, which couple to fermions~\footnote{Let me 
here refer to the simple case of subsect.~\ref{example2x2} by paying attention to the reader 
that in Table~\ref{Table VIII.} the two vacuum expectation values of each of the two scalar triplets, 
($\tilde{a}^{\tilde{N}_{L} \mp}, \tilde{a}^{\tilde{N}_{L} 3}$) and 
($\tilde{a}^{1 \mp}, \tilde{a}^{1 3}$), are expected to have the property ($\tilde{a}^{\tilde{N}_{L} + } \approx$
$\tilde{a}^{\tilde{N}_{L} -}$) and ($\tilde{a}^{1 +} \approx$ $ \tilde{a}^{1 -}$),  respectively,  
or at least very close to this. Then  superposition of the scalar fields, to 
which different families couple, might  differ a lot.} and depend 
on the quantum numbers $\alpha$ and  $k$, are in general   different 
from the superposition $\Phi^{\beta}$  (Eqs.~(\ref{scalarnew},\ref{scalaralphatimeevol})), which 
are the 
mass eigen states. Each family member $\alpha$ of each massive family $k$ couples in  general 
to different superposition of scalar fields. 

The two kinds of superposition are expressible with each other
\begin{eqnarray}
\label{PhiPhiPsi}
\Phi^{\alpha }_{f(o)\,k} &=& \sum_{\beta} \, D^{\alpha\,\beta}_{ k} \,\Phi^{\beta }\,. 
\end{eqnarray}
\section{Scalars which  bring masses to weak bosons}
\label{vectorbosonmasses}

In the {\it break} $I$ to the  vacuum of till this step two additional terms must be added (Eq.(\ref{vacuumI})), 
namely $\stackrel{78}{(-)} \ominus_{I} + \stackrel{78}{(+)} \oplus_{I}$, while $Q$ of this vacuum 
is equal to zero. 
We obtain for the mass term of the weak vector bosons, which interact with the scalar fields responsible 
for the appearance the {\it break} $I$, as follows
\begin{eqnarray}
\label{bosonmassdoubletItriplets}
 (\frac{1}{2})^2\, (g^1)^2 \,v_{I}^2(\frac{1}{(\cos \theta_{1})^2}\, Z^{Q'}_{m} Z^{Q' \,m} +  
 2\,W^{+}_{m} W^{- \,m} )\,.
\end{eqnarray}
To the vacuum expectation value $v_{I}$ all the scalar fields $\Phi^{I\,Ai}_{\mp}$ contribute. (These 
gauge vector bosons do not couple to the scalar doublets which cause the appearance of the {\it break} $II$.)


Let us conclude this section by recognizing that the contributions to masses of vector bosons 
are on the tree level determined by the coupling constants and the vacuum expectation values of these fields.

\section{Conclusions}
\label{discussionsandconclusions}

It is demonstrated (on the tree level) that according to the {\it spin-charge-family} theory -- 
which predicts the families of fermions and their charges, the gauge fields and several scalar fields --  
each family member  $\Psi^{\alpha \, k}$ ($\alpha = (u,d,\nu,e)$) of each family ($i=(1,2,3,4)$) 
couples to a different superposition of the scalar fields $\Phi^{\alpha}_{f(0)}$, Eq.~(\ref{Phipsi}), 
with the coupling constant proportional to its (fermion) mass (Eq.~(\ref{Phipsiex})). Each of these 
superposition differs from the superposition of the scalar fields which determine masses of gauge 
vector bosons (Eqs.~(\ref{bosonmassdoubletItriplets}, \ref{bosonmassdoubletIItriplets})), 
which further differ from the mass eigen states of scalars (Eq.~(\ref{scalaralphatimeevol})).

At each break 
the scalar fields gaining nonzero vacuum expectation values change the vacuum~(Eqs.(\ref{howStransform}, 
\ref{vacuumI})). At the electroweak break ({\it break}  $I$) the scalar fields, forbidden to behave 
as a vector in the $(\frac{1}{2}, \frac{1}{2})$ representation, behave as doublets (that is manifesting 
the fundamental representation) with respect to the weak $SU(2)$ group, while they behave like triplets 
with respect to the family groups 
($\tilde{\tau}^{1i}$ and $\tilde{N}^{i}_{L}$)  
and as singlets with respect to several $U(1)$  with the generators ($Y',Q',Q$). 
At the electroweak break 
the scalar fields, which determine fermion masses  on the 
tree level, take care of the differences in the masses among family members: i. Through  
differences in  their vacuum expectation values when they couple  to $(u^i,\nu^i)$ and 
when they  couple to $(d^i,e^i)$, and ii. Through differences in the eigen values of the operators 
($Y',Q',Q$) appearing with the corresponding scalar fields. 
Properties depend on the parameters, the values of which 
are in this paper not discussed. The fermion mass matrices manifest symmetries which limit strongly 
free parameters of the theory.

Since the mass term~(Eq.(\ref{factionM}))  
explains why the scalar fields behave like doublets
with respect to the weak $SU(2)$ group, while they keep the character of the adjoint representations 
with respect to the family groups, the {\it standard model} can really be interpreted 
as a low energy effective manifestation of the {\it spin-charge-family} theory.  

Let me say that I assumed (make a choice among the possibilities) such breaks of symmetries of a 
simple starting action which lead to observable phenomena. It is, however, far from saying that 
I move the assumptions of the {\it standard model} to next step, since many assumptions of 
the {\it standard model} get explanations through the  {\it spin-charge-family} theory.

Let me conclude the paper with the predictions of the {\it spin-charge-family} theory: 
Observations of the scalar fields at  the LHC and other experiments are expected to differ  from 
the predictions of the {\it standard model}, although so far the experimental data have shown 
no disagreement with the {\it standard model} predictions. Several scalar fields, predicted by the theory,
will show up. The fourth family will sooner or later be observed and it might be that the observed ten-jet 
event is caused by the fourth family members.  There will be no supersymmetric partners.

A systematic study of predictions, which would take into account the loop corrections 
of the {\it spin-charge-family} theory  is needed and it is  
in progress~\cite{GNBled2012,GNinprogress}. The calculations~\cite{GNBled2012}, in particular,  manifest
that, if  symmetries of the $4\times 4$ mass matrices required by the theory are respected 
in all orders in loop corrections,   the observed masses and mixing matrices of the 
lower three families, when included into $4\times 4$ lead to mass matrices quite close (within a factor 
of $3$) to the democratic matrices, equivalently for quarks and leptons.

\appendix*

\section{The technique for representing spinors~\cite{NF,norma,norma93,hn0203}}
\label{cliffordfamilies} 

The technique~\cite{NF,norma93,hn0203,norma} can be used to construct a spinor basis for any dimension $d$
and any signature in an easy and transparent way. Equipped with the graphic presentation of basic states,  
the technique offers an elegant way to see all the quantum numbers of states with respect to the two 
Lorentz groups with the infinitesimal generators of the groups $S^{ab}$ and $\tilde{S}^{ab}$, 
as well as transformation properties of the states under any Clifford algebra object $\gamma^a$ 
and $\tilde{\gamma}^a$, $ \{ \gamma^a, \gamma^b\}_{+} = 2\eta^{ab}\,$,    
$\{ \tilde{\gamma}^a, \tilde{\gamma}^b\}_{+}= 2\eta^{ab}\,$, 
$\{ \gamma^a, \tilde{\gamma}^b\}_{+} = 0\,$, for any $d$, even or odd.  

Since the Clifford algebra objects $S^{ab}= (i/4) (\gamma^a \gamma^b - \gamma^b \gamma^a)\,$ and 
$\tilde{S}^{ab}= (i/4) (\tilde{\gamma}^a \tilde{\gamma}^b 
- \tilde{\gamma}^b \tilde{\gamma}^a)\,$ close the algebra of the Lorentz group, while 
$\{S^{ab}, \tilde{S}^{cd}\}_{-}= 0\,$, 
$S^{ab}$ and $\tilde{S}^{ab}$ form the equivalent representations to each other.
If $S^{ab}$ are used to determine spinor representations in $d$ dimensional space, and after the 
break of symmetries, the spin and the charges in $d= (1+3)$, can $\tilde{S}^{ab}$ be used to describe 
families of spinors. 

To make the technique simple  the graphic presentation of nilpotents and projectors was introduced~\cite{hn0203}. 
For even $d$ we have
\begin{eqnarray}
\stackrel{ab}{(k)}:&=& 
\frac{1}{2}(\gamma^a + \frac{\eta^{aa}}{ik} \gamma^b)\,,\quad \quad
\stackrel{ab}{[k]}:=
\frac{1}{2}(1+ \frac{i}{k} \gamma^a \gamma^b)\,,
\label{signature}
\end{eqnarray}
with the properties $k^2 = \eta^{aa} \eta^{bb}$ and 
\begin{eqnarray}
        S^{ab}\, \stackrel{ab}{(k)}= \frac{1}{2}\,k\, \stackrel{ab}{(k)}\,,\quad  
        S^{ab}\, \stackrel{ab}{[k]}= \frac{1}{2}\,k \,\stackrel{ab}{[k]}\,,\quad 
\tilde{S}^{ab}\, \stackrel{ab}{(k)}= \frac{1}{2}\,k \,\stackrel{ab}{(k)}\,,\quad  
\tilde{S}^{ab}\, \stackrel{ab}{[k]}=-\frac{1}{2}\,k \,\stackrel{ab}{[k]}\,.
\label{grapheigen}
\end{eqnarray}
One recognizes 
that $\gamma^a$ transform  $\stackrel{ab}{(k)}$ into  $\stackrel{ab}{[-k]}$, never to 
$\stackrel{ab}{[k]}$, while $\tilde{\gamma}^a$ transform  $\stackrel{ab}{(k)}$ into 
$\stackrel{ab}{[k]}$, never to $\stackrel{ab}{[-k]}$ 
\begin{eqnarray}
&&\gamma^a \stackrel{ab}{(k)}= \eta^{aa}\stackrel{ab}{[-k]},\; 
\gamma^b \stackrel{ab}{(k)}= -ik \stackrel{ab}{[-k]}, \; 
\gamma^a \stackrel{ab}{[k]}= \stackrel{ab}{(-k)},\; 
\gamma^b \stackrel{ab}{[k]}= -ik \eta^{aa} \stackrel{ab}{(-k)}\,,\nonumber\\
&&\tilde{\gamma^a} \stackrel{ab}{(k)} = - i\eta^{aa}\stackrel{ab}{[k]},\;
\tilde{\gamma^b} \stackrel{ab}{(k)} =  - k \stackrel{ab}{[k]}, \;
\tilde{\gamma^a} \stackrel{ab}{[k]} =  \;\;i\stackrel{ab}{(k)},\; 
\tilde{\gamma^b} \stackrel{ab}{[k]} =  -k \eta^{aa} \stackrel{ab}{(k)}\,. 
\label{snmb:gammatildegamma}
\end{eqnarray}

Let us add some useful relations
\begin{eqnarray}
\stackrel{ab}{(k)}\stackrel{ab}{(k)}& =& 0\,, \quad \quad \stackrel{ab}{(k)}\stackrel{ab}{(-k)}
= \eta^{aa}  \stackrel{ab}{[k]}\,, \quad  
\stackrel{ab}{[k]}\stackrel{ab}{[k]} = \stackrel{ab}{[k]}\,, \quad \quad
\stackrel{ab}{[k]}\stackrel{ab}{[-k]}= 0\,, 
 \nonumber\\
\stackrel{ab}{(k)}\stackrel{ab}{[k]}& =& 0\,,\quad \quad \quad \stackrel{ab}{[k]}\stackrel{ab}{(k)}
=  \stackrel{ab}{(k)}\,, \quad \quad \quad \stackrel{ab}{(-k)}\stackrel{ab}{[k]}=
 \stackrel{ab}{(-k)}\,,
\quad \quad \stackrel{ab}{[k]}\stackrel{ab}{(-k)} =0 \,.  
\label{graphbinoms}
\end{eqnarray}
Defining
\begin{eqnarray}
\stackrel{ab}{\tilde{(\pm i)}} = 
\frac{1}{2} \, (\tilde{\gamma}^a \mp \tilde{\gamma}^b)\,, \quad
\stackrel{ab}{\tilde{(\pm 1)}} = 
\frac{1}{2} \, (\tilde{\gamma}^a \pm i\tilde{\gamma}^b)\,, 
\label{deftildefun}
\end{eqnarray}
it follows 
\begin{eqnarray}
\stackrel{ab}{\tilde{( k)}} \, \stackrel{ab}{(k)}& =& 0\,, 
\quad \;
\stackrel{ab}{\tilde{(-k)}} \, \stackrel{ab}{(k)} = -i \eta^{aa}\,  \stackrel{ab}{[k]}\,,
\quad\;
\stackrel{ab}{\tilde{( k)}} \, \stackrel{ab}{[k]} = i\, \stackrel{ab}{(k)}\,,
\quad\;
\stackrel{ab}{\tilde{( k)}}\, \stackrel{ab}{[-k]} = 0\,.
\label{graphbinomsfamilies}
\end{eqnarray}
We define the vacuum $|\psi_0>$ so that 
$< \;\stackrel{ab}{(k)}^{\dagger} \stackrel{ab}{(k)}\; > = 1\,$ and $< \;\stackrel{ab}{[k]}^{\dagger}
 \stackrel{ab}{[k]}\; > = 1\,$.

Making a choice of the Cartan subalgebra set of the algebra $S^{ab}$ and $\tilde{S}^{ab}$  
$(S^{03}, S^{12}, S^{56}, S^{78}, \cdots)$ and $(\tilde{S}^{03}, \tilde{S}^{12}, \tilde{S}^{56}, 
\tilde{S}^{78},\cdots )$ 
an eigen state of all the members of the Cartan  subalgebra, representing a weak chargeless  
$u_{R}$-quark with spin up, hyper charge ($2/3$) and  colour ($1/2\,,1/(2\sqrt{3})$), for example, 
can be written as 
%
$ \stackrel{03}{(+i)}\stackrel{12}{(+)}|\stackrel{56}{(+)}\stackrel{78}{(+)}
||\stackrel{9 \;10}{(+)}\stackrel{11\;12}{(-)}\stackrel{13\;14}{(-)} |\psi \rangle$  $= 
\frac{1}{2^7} 
(\gamma^0 -\gamma^3)(\gamma^1 +i \gamma^2)$ $| (\gamma^5 + i\gamma^6)(\gamma^7 +i \gamma^8)$ $||
(\gamma^9 +i\gamma^{10})(\gamma^{11} -i \gamma^{12})(\gamma^{13}-i\gamma^{14})
|\psi \rangle \,$.
%
This state is an eigen state of all $S^{ab}$ and $\tilde{S}^{ab}$ which are members of the Cartan 
subalgebra. The definition of the  charges can be found in the ref.~\cite{pikanorma,NF}.

In Table~\ref{Table I.} the eightplet of quarks of a particular colour charge ($\tau^{33}=1/2$, 
 $\tau^{38}=1/(2\sqrt{3})$) and the $U(1)_{II}$ charge ($\tau^{4}=1/6$) is presented in our 
 technique~\cite{norma93,hn0203}, as products of nilpotents and projectors. 
 \begin{table}
 \begin{center}
 \begin{tabular}{|r|c||c||c|c|c|r|r|}
 \hline
 i&$$&$|^a\psi_i>$&$\Gamma^{(1,3)}$&$ S^{12}$&
 $\tau^{13}$&$Y$&$Q$\\
 \hline\hline
 && ${\rm Octet} \,\,{\rm of \; quarks}$&&&&&\\
 \hline\hline
 1&$ u_{R}^{c1}$&$ \stackrel{03}{(+i)}\,\stackrel{12}{(+)}|
 \stackrel{56}{(+)}\,\stackrel{78}{(+)}
 ||\stackrel{9 \;10}{(+)}\;\;\stackrel{11\;12}{[-]}\;\;\stackrel{13\;14}{[-]} $
 &1&$\frac{1}{2}$&0&$\frac{2}{3}$&$\frac{2}{3}$\\
 \hline 
 2&$u_{R}^{c1}$&$\stackrel{03}{[-i]}\,\stackrel{12}{[-]}|\stackrel{56}{(+)}\,\stackrel{78}{(+)}
 ||\stackrel{9 \;10}{(+)}\;\;\stackrel{11\;12}{[-]}\;\;\stackrel{13\;14}{[-]}$
 &1&$-\frac{1}{2}$&0&$\frac{2}{3}$&$\frac{2}{3}$\\
 \hline
 3&$d_{R}^{c1}$&$\stackrel{03}{(+i)}\,\stackrel{12}{(+)}|\stackrel{56}{[-]}\,\stackrel{78}{[-]}
 ||\stackrel{9 \;10}{(+)}\;\;\stackrel{11\;12}{[-]}\;\;\stackrel{13\;14}{[-]}$
 &1&$\frac{1}{2}$&0&$-\frac{1}{3}$&$-\frac{1}{3}$\\
 \hline 
 4&$ d_{R}^{c1} $&$\stackrel{03}{[-i]}\,\stackrel{12}{[-]}|
 \stackrel{56}{[-]}\,\stackrel{78}{[-]}
 ||\stackrel{9 \;10}{(+)}\;\;\stackrel{11\;12}{[-]}\;\;\stackrel{13\;14}{[-]} $
 &1&$-\frac{1}{2}$&0&$-\frac{1}{3}$&$-\frac{1}{3}$\\
 \hline
 5&$d_{L}^{c1}$&$\stackrel{03}{[-i]}\,\stackrel{12}{(+)}|\stackrel{56}{[-]}\,\stackrel{78}{(+)}
 ||\stackrel{9 \;10}{(+)}\;\;\stackrel{11\;12}{[-]}\;\;\stackrel{13\;14}{[-]}$
 &-1&$\frac{1}{2}$&$-\frac{1}{2}$&$\frac{1}{6}$&$-\frac{1}{3}$\\
 \hline
 6&$d_{L}^{c1} $&$\stackrel{03}{(+i)}\,\stackrel{12}{[-]}|
 \stackrel{56}{[-]}\,\stackrel{78}{(+)}
 ||\stackrel{9 \;10}{(+)}\;\;\stackrel{11\;12}{[-]}\;\;\stackrel{13\;14}{[-]} $
 &-1&$-\frac{1}{2}$&$-\frac{1}{2}$&$\frac{1}{6}$&$-\frac{1}{3}$\\
 \hline
 7&$ u_{L}^{c1}$&$\stackrel{03}{[-i]}\,\stackrel{12}{(+)}|
 \stackrel{56}{(+)}\,\stackrel{78}{[-]}
 ||\stackrel{9 \;10}{(+)}\;\;\stackrel{11\;12}{[-]}\;\;\stackrel{13\;14}{[-]}$
 &-1&$\frac{1}{2}$&$\frac{1}{2}$&$\frac{1}{6}$&$\frac{2}{3}$\\
 \hline
 8&$u_{L}^{c1}$&$\stackrel{03}{(+i)}\,\stackrel{12}{[-]}|\stackrel{56}{(+)}\,\stackrel{78}{[-]}
 ||\stackrel{9 \;10}{(+)}\;\;\stackrel{11\;12}{[-]}\;\;\stackrel{13\;14}{[-]}$
 &-1&$-\frac{1}{2}$&$\frac{1}{2}$&$\frac{1}{6}$&$\frac{2}{3}$\\
 \hline\hline
 \end{tabular}
 \end{center}
 \caption{\label{Table I.} The 8-plet of quarks - the members of $SO(7,1)$ subgroup of the 
 group $SO(13,1)$, belonging to one Weyl 
 spinor representation of  $SO(13,1)$ is presented in the technique~\cite{hn0203}. 
 It contains the left handed weak charged quarks and the right handed weak chargeless quarks 
 of a particular 
 colour $(1/2,1/(2\sqrt{3}))$. Here  $\Gamma^{(1,3)}$ defines the handedness in $(1+3)$ space, 
 $ S^{12}$ defines the ordinary spin, 
 $\tau^{13}$ defines the third component of the weak charge,  
 $Y$ is the hyper charge,  
 $Q= Y + \tau^{13}$ is the 
 electromagnetic charge. The vacuum state $|\psi_0>$, on which the nilpotents and 
 projectors operate, is not shown. The basis is the massless one. One easily sees that 
 $\gamma^0 \,\stackrel{78}{(-)}$ transforms the first line ($u_{R}^{c1}$) into the seventh one ($u_{L}^{c1}$), 
 and $\gamma^0 \,\stackrel{78}{(+)}$ transforms the third line ($d_{R}^{c1}$) into the fifth one ($d_{L}^{c1}$).
 }
 \end{table}
The operators $ \tilde{S}^{ab}$ 
generate families from the starting $u_R$ quark, transforming $u_R$ quark from Table~(\ref{Table VIII.}) 
to the $u_R$ of another family,  keeping all the properties with respect to $S^{ab}$ unchanged. 
The eight families of the first 
 member of the eightplet of quarks from Table~\ref{Table I.}, for example, that is of the right 
 handed $u^{c1}_{R}$-quark  with spin $\frac{1}{2}$,  are presented in the left column of 
 Table~\ref{Table III.}. 
 The eight-plet of  the corresponding 
 right handed neutrinos with spin up is  presented  
 in the right column of the same table. All the other members of any of the eight families of quarks or 
 leptons follow  from any member of a particular family by the application of  the 
 operators  $S^{ab}$ 
 on this particular member.  
 \begin{table}
 \begin{center}
 \begin{tabular}{|r||c||c||c||c||}
 \hline
 $I_R$ & $u_{R}^{c1}$&
  $ \stackrel{03}{(+i)}\,\stackrel{12}{[+]}|\stackrel{56}{[+]}\,\stackrel{78}{(+)}||
  \stackrel{9 \;10}{(+)}\;\;\stackrel{11\;12}{[-]}\;\;\stackrel{13\;14}{[-]}$ & 
  $\nu_{R}$&
  $ \stackrel{03}{[+i]}\,\stackrel{12}{(+)}|\stackrel{56}{[+]}\,\stackrel{78}{(+)}|| 
  \stackrel{9 \;10}{(+)}\;\;\stackrel{11\;12}{(+)}\;\;\stackrel{13\;14}{(+)}$ 
 \\
 \hline
  $II_R$ & $u_{R}^{c1}$&
  $ \stackrel{03}{[+i]}\,\stackrel{12}{(+)}|\stackrel{56}{[+]}\,\stackrel{78}{(+)}||
  \stackrel{9 \;10}{(+)}\;\;\stackrel{11\;12}{[-]}\;\;\stackrel{13\;14}{[-]}$ & 
  $\nu_{R}$&
  $ \stackrel{03}{(+i)}\,\stackrel{12}{[+]}|\stackrel{56}{(+)}\,\stackrel{78}{[+]}||
  \stackrel{9 \;10}{(+)}\;\;\stackrel{11\;12}{(+)}\;\;\stackrel{13\;14}{(+)}$ 
 \\
 \hline
 $III_R$ & $u_{R}^{c1}$&
 $ \stackrel{03}{(+i)}\,\stackrel{12}{[+]}|\stackrel{56}{(+)}\,\stackrel{78}{[+]}||
 \stackrel{9 \;10}{(+)}\;\;\stackrel{11\;12}{[-]}\;\;\stackrel{13\;14}{[-]}$ & 
 $\nu_{R}$&
 $ \stackrel{03}{(+i)}\,\stackrel{12}{[+]}|\stackrel{56}{[+]}\,\stackrel{78}{(+)}||
 \stackrel{9 \;10}{(+)}\;\;\stackrel{11\;12}{(+)}\;\;\stackrel{13\;14}{(+)}$ 
 \\
 \hline
 $IV_R$ & $u_{R}^{c1}$&
  $ \stackrel{03}{[+i]}\,\stackrel{12}{(+)}|\stackrel{56}{(+)}\,\stackrel{78}{[+]}||
  \stackrel{9 \;10}{(+)}\:\; \stackrel{11\;12}{[-]}\;\;\stackrel{13\;14}{[-]}$ & 
  $\nu_{R}$&
  $ \stackrel{03}{[+i]}\,\stackrel{12}{(+)}|\stackrel{56}{(+)}\,\stackrel{78}{[+]}||
  \stackrel{9 \;10}{(+)}\;\;\stackrel{11\;12}{(+)}\;\;\stackrel{13\;14}{(+)}$ 
 \\
 \hline 
 \end{tabular}
 \end{center}
 \caption{\label{Table III.} Four  families of the right handed $u^{c1}_R$ quark with spin 
 $\frac{1}{2}$, the colour charge (${}^{c1}$ $=(\tau^{33}=1/2$, $\tau^{38}=1/(2\sqrt{3}))$, and of 
 the colourless right handed neutrino $\nu_R$ of spin $\frac{1}{2}$ are presented in the 
 left and in the right column, respectively. All the families follow from the starting one by the 
 application of the operators  $\tilde{S}^{ab}$, $a,b \in\{0,1,2,\cdots, 8\}$. 
The  generators $S^{ab}$, $a,b \in\{0,1,2,\cdots, 8\}$  transform 
equivalently the right handed   neutrino $\nu_R$ of  spin $\frac{1}{2}$ to all the colourless 
members of the same family. 
}
 \end{table}

Let us add below some useful relations~\cite{pikanorma} 
\begin{eqnarray}
\label{plusminus}
N^{\pm}_{+}         &=& N^{1}_{+} \pm i \,N^{2}_{+} = 
 - \stackrel{03}{(\mp i)} \stackrel{12}{(\pm )}\,, \quad N^{\pm}_{-}= N^{1}_{-} \pm i\,N^{2}_{-} = 
  \stackrel{03}{(\pm i)} \stackrel{12}{(\pm )}\,,\nonumber\\
\tilde{N}^{\pm}_{+} &=& - \stackrel{03}{\tilde{(\mp i)}} \stackrel{12}{\tilde{(\pm )}}\,, \quad 
\tilde{N}^{\pm}_{-}= 
  \stackrel{03} {\tilde{(\pm i)}} \stackrel{12} {\tilde{(\pm )}}\,,\nonumber\\ 
\tau^{1\pm}         &=& (\mp)\, \stackrel{56}{(\pm )} \stackrel{78}{(\mp )} \,, \quad   
\tau^{2\mp}=            (\mp)\, \stackrel{56}{(\mp )} \stackrel{78}{(\mp )} \,,\nonumber\\ 
\tilde{\tau}^{1\pm} &=& (\mp)\, \stackrel{56}{\tilde{(\pm )}} \stackrel{78}{\tilde{(\mp )}}\,,\quad   
\tilde{\tau}^{2\mp}= (\mp)\, \stackrel{56}{\tilde{(\mp )}} \stackrel{78}{\tilde{(\mp )}}\,,\nonumber\\
\tau^{Ai}&=& \sum_{ab}\, C^{Ai}_{ab}\, S^{ab}\,,\quad \tilde{\tau}^{Ai}= \sum_{ab}\, C^{Ai}_{ab}\, 
\tilde{S}^{ab}\,.
\end{eqnarray}
 \section{Scalars contributing to the {\it  break} II}
 \label{breakII}

 In section~\ref{SCFT} scalar fields $\Phi^{I\,Ai}$, contributing to masses of the lower four 
 families (among which are  the three so far observed families), are discussed. Here we discuss 
 properties of scalar fields $\Phi^{II\,Ai}$, contributing to masses  of the upper four families.

 Let $\Phi^{II\,Ai}$ stay for all the scalar fields contributing to massess of the upper four 
 families and to the masses of  the gauge vector bosons $\vec{A}^{2}_{m}$, $m=0,1,2,3$
 \begin{eqnarray}
 \label{allscalarsII}
 \Phi^{II\,Ai} & \equiv&  \Phi^{II\, Ai}_{\mp}\,, \quad  
 \Phi^{II\, Ai}_{\mp}= (\vec{\tilde{A}}^{2}_{\mp}\,,\vec{\tilde{A}}^{\tilde{N}_R}_{\mp})\,,
  \nonumber\\
 \Phi^{Ai}_{\mp}= (\Phi^{Ai}_{7} \pm i \Phi^{Ai}_{8})\,, &&\quad A_{II}= \{\tilde{2}, \tilde{N}_R\}\,.
 \end{eqnarray}
 We  choose a renormalizable effective potential $V(\Phi^{II,Ai})  $ for the (assumed to be) real 
 scalar fields $\Phi^{II,Ai}$ (Eq.~(\ref{allscalars})), which couple among themselves 
 %
 %
 \begin{eqnarray}
 \label{veffII}
 V(\Phi^{II\,Ai})   &=& \sum_{A,i}\{\, -\frac{1}{2} \,  (m^{II}_{Ai})^2 (\Phi^{II\,Ai})^2 + 
 \frac{1}{4}\, \sum_{ B,
 j}\, \lambda^{II\,Ai\,Bj}\, (\Phi^{II\,Ai})^2 \, (\Phi^{II\,Bj})^2\}\,,
 \end{eqnarray}
with $\lambda^{Ai\,Bj}=\lambda^{Bj\, Ai}$.

Scalar fields couple to the gauge bosons  at the {\it break } $II$ 
according to the Lagrange function  ${\mathcal L}_{s\,I}$
\begin{eqnarray}
\label{scalarLagrange0II}
{\mathcal L}_{s\,II} &=& \sum_{A,i}\,\,(p_{0m} \Phi^{II\,Ai}_{})^{\dagger}(p_{0}{}^{m}\, \Phi^{II \,Ai}_{})
- V(\Phi^{II\,Ai})\,,
\nonumber\\
p_{0m} &=& p_{m} - \{ g^{4}\ \,\tau^{4}\,A^{4}_{m} +  g^{2}\, \vec{\tau}^2\, \vec{A}^{2}_{m} \}\,. 
\end{eqnarray}

The vacuum state can be expressed before the {\it break} $II$ (due to our 
technique~\cite{norma93,pikanorma,hn0203}) as an identity with respect to $SO(4)$ 
(on which products of  nilpotents and projectors, see Table~\ref{Table I.} of 
appendix~\ref{cliffordfamilies}, representing fermion states  
apply). To this vacuum  a new kind of 
terms, manifesting as a doublet with respect to the weak charge and as triplets with 
respect to the charges of $\tilde{S}^{ab}$ origin, must be added
\begin{eqnarray}
\label{vacuumII}
\stackrel{78}{(-)} \ominus_{II}&=&
\stackrel{78}{(-)}
Ts_{\tilde{N}_R} 
|(\stackrel{56}{[-]} \stackrel{78}{(+)})
\,Td_{(-)\tilde{\tau}^2}
||\stackrel{9\,10}{[-]}\stackrel{11\,12}{[-]} \stackrel{13\,14}{[-]}\,,
\nonumber\\
\stackrel{78}{(+)} \oplus_{II}&=&
\stackrel{78}{(+)}
Ts_{\tilde{N}_R} 
 | 
(\stackrel{56}{[+]} \stackrel{78}{(-)})
\,Td_{(+) \tilde{\tau}^2}
||\stackrel{9\,10}{[+]}\stackrel{11\,12}{[+]} \stackrel{13\,14}{[+]}
\,.
\end{eqnarray}
Here $Ts_{\tilde{N}_R} $ denotes a triplet with respect to the operators $\vec{\tilde{N}}_{R}$ and 
a singlet with respect to $\vec{N}_{R}$, while $(\stackrel{56}{[+]} \stackrel{78}{(-)})
\,Td_{(\mp) \tilde{\tau}^2}$ are the two triplets with respect to $\vec{\tilde{\tau}}^{2}$ which are 
doublets with respect 
to $\vec{\tau}^2$. It is not difficult, using our technique,  to define such a vacuum as  products of two 
spinor representations. One can check that the products of projectors and nilpotents defining 
the members of one family and correspondingly of all the families, when applied on these two additional 
contributions to the vacuum, gives zero.
One finds
\begin{eqnarray}
\label{propvacuumII}
\tau^{2+} \tau^{2-}\,\stackrel{78}{(+)} \oplus_{II} &=& \stackrel{78}{(+)} \oplus_{II}\,,\quad \quad 
\tau^{2-} \tau^{2+}\,\stackrel{78}{(-)} \ominus_{II} = \stackrel{78}{(-)} \ominus_{II}\,,\nonumber\\
Y \,\stackrel{78}{(+)} \oplus_{II} &=& 0 = Y \,\stackrel{78}{(-)} \ominus_{II}\,,\quad 
Q \,\stackrel{78}{(+)} \oplus_{II}  =  0 = Q \,\stackrel{78}{(-)} \ominus_{II}\,, \nonumber\\
Y' \,\stackrel{78}{(+)} \oplus_{II} &=& \frac{1}{2\cos^2 \theta_2}\,,\quad 
Y' \,\stackrel{78}{(-)} \ominus_{II} = -\frac{1}{2\cos^2 \theta_2}\,,\nonumber\\
(\tau^{1+}, \tau^{1-}, \tau^{1\,3}) \stackrel{78}{(+)} \oplus_{II} &=& 0= 
(\tau^{1+}, \tau^{1-}, \tau^{1\,3}) \stackrel{78}{(-)} \ominus_{II}\,.
\end{eqnarray}
We see that scalar fields $\phi^{II\,Ai}_{\mp}$~(Eq.\ref{allscalars}) bring masses to the gauge 
vector bosons $A^{2\pm}_m$ and $A^{Y'}_{m}= \cos \theta_2 \,A^{23}_m - \sin \theta_2\, A^{4}_{m}$, while 
$A^{Y}_{m}= \sin \theta_2 \,A^{23}_m + \cos \theta_2 \,A^{4}_{m}$  stays 
massless, provided that $\frac{g^2}{g^4} \,\tan \theta_2=1$. Also the weak gauge vectors 
$\vec{A}^{1}_m$ stay massless.

At each break the vacuum changes (Eq.~(\ref{vacuumII})). At  the {\it break} $II$ it changes to 
($I + \stackrel{78}{(-)} \ominus_{II} + \stackrel{78}{(+)} \oplus_{II}$).   $Q$ and 
$Y$ of this vacuum are $0$, and due to that fact that the expectation values of $\vec{\tilde{\tau}}^2$ and 
$\vec{\tilde{N}}_{R}$ are not influenced by $\vec{\tau}^2$ or $Y'$, one easily finds for the mass term of the 
$SU(2)_{II}$ and $U(1)_{II}$ vector gauge fields (see Eq.~(\ref{propvacuumII}) and the text below this equation)
after the {\it break} $II$
%
\begin{eqnarray}
\label{bosonmassdoubletIItriplets}
 (\frac{1}{2})^2\, (g^2)^2 \,v_{II}^2(\frac{1}{(\cos \theta_{2})^2}\, A^{Y'}_{m} A^{Y' \,m} +  
 2\,A^{2+}_{m} A^{2- \,m} )\,.
\end{eqnarray}
To the vacuum expectation value $v_{II}$ all the scalar fields $\Phi^{II\,Ai}_{\mp}$ contribute. 

\acknowledgements  
The research was financed by the Slovenian Research Agency, project no. P1-0188.

\end{document}